\documentclass[12pt]{article}
\usepackage{epsfig}
\usepackage{amssymb}
\usepackage{amsbsy}
\hoffset= - 1.0in         % switch off for draft style

\catcode`\@=11
\def\marginnote#1{}

\newcount\hour
\newcount\minute
\newtoks\amorpm
\hour=\time\divide\hour by60
\minute=\time{\multiply\hour by60 \global\advance\minute by-\hour}
\edef\standardtime{{\ifnum\hour<12 \global\amorpm={am}%
        \else\global\amorpm={pm}\advance\hour by-12 \fi
        \ifnum\hour=0 \hour=12 \fi
        \number\hour:\ifnum\minute<10 0\fi\number\minute\the\amorpm}}
\edef\militarytime{\number\hour:\ifnum\minute<10 0\fi\number\minute}

\def\draftlabel#1{{\@bsphack\if@filesw {\let\thepage\relax
   \xdef\@gtempa{\write\@auxout{\string
      \newlabel{#1}{{\@currentlabel}{\thepage}}}}}\@gtempa
   \if@nobreak \ifvmode\nobreak\fi\fi\fi\@esphack}
        \gdef\@eqnlabel{#1}}
\def\@eqnlabel{}
\def\@vacuum{}
\def\draftmarginnote#1{\marginpar{\raggedright\scriptsize\tt#1}}

\def\draft{\oddsidemargin -.5truein
        \def\@oddfoot{\sl preliminary draft \hfil
        \rm\thepage\hfil\sl\today\quad\militarytime}
        \let\@evenfoot\@oddfoot \overfullrule 3pt
        \let\label=\draftlabel
        \let\marginnote=\draftmarginnote
   \def\@eqnnum{(\theequation)\rlap{\kern\marginparsep\tt\@eqnlabel}%
\global\let\@eqnlabel\@vacuum}  }
\makeatletter
\newdimen\normalarrayskip              % skip between lines
\newdimen\minarrayskip                 % minimal skip between lines
\normalarrayskip\baselineskip
\minarrayskip\jot
\newif\ifold             \oldtrue            \def\new{\oldfalse}
\def\arraymode{\ifold\relax\else\displaystyle\fi}%mode of array entries
\def\eqnumphantom{\phantom{(\theequation)}} % right phantom in eqnarray
\def\@arrayskip{\ifold\baselineskip\z@\lineskip\z@
     \else
     \baselineskip\minarrayskip\lineskip2\minarrayskip\fi}
\def\@arrayclassz{\ifcase \@lastchclass \@acolampacol \or
\@ampacol \or \or \or \@addamp \or
   \@acolampacol \or \@firstampfalse \@acol \fi
\edef\@preamble{\@preamble
  \ifcase \@chnum
     \hfil$\relax\arraymode\@sharp$\hfil
     \or $\relax\arraymode\@sharp$\hfil
     \or \hfil$\relax\arraymode\@sharp$\fi}}
\def\@array[#1]#2{\setbox\@arstrutbox=\hbox{\vrule
     height\arraystretch \ht\strutbox
     depth\arraystretch \dp\strutbox
     width\z@}\@mkpream{#2}\edef\@preamble{\halign
\noexpand\@halignto
\bgroup \tabskip\z@ \@arstrut \@preamble \tabskip\z@ \cr}%
\let\@startpbox\@@startpbox \let\@endpbox\@@endpbox
  \if #1t\vtop \else \if#1b\vbox \else \vcenter \fi\fi
  \bgroup \let\par\relax
  \let\@sharp##\let\protect\relax
  \@arrayskip\@preamble}
\def\eqnarray{\stepcounter{equation}%
              \let\@currentlabel=\theequation
              \global\@eqnswtrue
              \global\@eqcnt\z@
              \tabskip\@centering
              \let\\=\@eqncr
 \halign to \displaywidth\bgroup
    \eqnumphantom\@eqnsel\hskip\@centering
    $\displaystyle \tabskip\z@ {##}$%
    \global\@eqcnt\@ne \hskip 2\arraycolsep
         $\displaystyle\arraymode{##}$\hfil
    \global\@eqcnt\tw@ \hskip 2\arraycolsep
         $\displaystyle\tabskip\z@{##}$\hfil
         \tabskip\@centering
    &{##}\tabskip\z@\cr}
\def\R{\mathbb{R}}
\def\CN{{\cal N}}

\def\e{{\,\rm e}\,}
\def\d{\partial}
\def\p#1#2{\partial #1\over\partial #2}

\def\bea{\begin{eqnarray}}
\def\eea{\end{eqnarray}}

\def\be{\begin{equation}}
\def\ee{\end{equation}}
\def\ba{\beq\new\begin{array}{c}}
\def\ea{\end{array}\eeq}
\def\stackreb#1#2{\mathrel{\mathop{#2}\limits_{#1}}}
\def\Tr{\,{\rm Tr}\,}\def\tr{\,{\rm tr}\,}

\def\Im{{\rm Im}}

\def\2{{1\over 2}}
\def\N2{${\cal N}=2$}
\def\4N{${\cal N}=4$}
\def\1N{${\cal N}=1$}

\def\eps{\epsilon}

\def\ha{{1\over 2}}

\def\pint{{-\!\!\!\!\!\!\int}}
\newcommand{\rf}[1]{(\ref{#1})}

\def\D{\Delta}
\def\p{\partial}
\def\({\left(}
\def\){\right)}
\def\hf{{1\over 2}}

\def\s{\sigma}

\def\ads{$AdS_5\times S^5$}
\def\der{D_+}
\def\sm{{\bf s}}

\def\sT{\sqrt{T}}

\newcommand{\sectiono}[1]{\section{#1}\setcounter{equation}{0}}
\renewcommand{\theequation}{\thesection.\arabic{equation}}
\begin{document}

\thispagestyle{empty}
\begin{flushright}%\footnotesize\tt
%hep-th/0410105\\
LPTENS-04/41\\
MPG/ITEP-42/04\\
UUITP-21/04\\
\end{flushright}
\vspace{.5cm}
\setcounter{footnote}{0}
\begin{center}
{\Large{\bf Classical/quantum integrability \\ \vskip 3mm in
non-compact sector of AdS/CFT \par}
   }\vspace{4mm}
{\large\rm V.A.~Kazakov$^{a,}$\footnote{ Membre de
l'Institut Universitaire
    de France}
and K. Zarembo$^{b,}$\footnote{Also at ITEP, Moscow, 117259
Bol. Cheremushkinskaya 25, Russia}\\[7mm]
\large\it ${}^a$ Laboratoire de Physique Th\'eorique de
l'Ecole Normale
Sup\'erieure et l'Universit\'e Paris-6,\\
Paris, Cedex 75231, France}\\[2mm]
{\large\it ${}^b$ Institutionen f\"or Teoretisk Fysik,
Uppsala Universitet\\
Box 803, SE-751 08 Uppsala,  Sweden}\\[10mm]

%\vskip 1cm

{\tt\noindent kazakov@physique.ens.fr\\
\indent\ \  konstantin.zarembo@teorfys.uu.se}

\vskip 1cm

%\newpage
{\sc Abstract}\\[2mm]
\end{center}
\noindent We discuss  non-compact $SL(2,\R)$ sectors in N=4 SYM and
in AdS string theory and compare their integrable structures. We
formulate and solve the Riemann-Hilbert problem for the finite gap
solutions of the classical sigma model and show that at one loop it
is identical to the classical limit of Bethe equations of the spin
(-1/2) chain for the dilatation operator of SYM.

%We show that classical solutions of the sigma-model are encoded in
%an integral equation of the Bethe type which agrees with the Bethe
%equations in SYM at one loop.
%We found some indications for
%discrepancies at the two-loop level.

\newpage
\setcounter{page}{1}
\renewcommand{\thefootnote}{\arabic{footnote}}
\setcounter{footnote}{0}

%%%%%%%%%%%%%%%%%%%%%%%%%%%%%%%%%%%%%%%%%%%%
    \sectiono{Introduction}
%%%%%%%%%%%%%%%%%%%%%%%%%%%%%%%%%%%%%%%%%%%%

The  semiclassical limit of the AdS/CFT correspondence
\cite{Berenstein:2002jq,Gubser:2002tv} reveals new symmetries which
are likely to play an important role in the poorly understood
quantum regime of the duality. The semiclassical approximation is
accurate for  states (closed string states in AdS or local operators
in CFT) whose quantum numbers are large.
%Since individual string
%states are dual to local operators,
While string theory certainly simplifies in this limit, the
necessity to consider operators with large quantum numbers is a
complication rather than simplification in the field theory. Such
operators contain many constituent fields, are highly degenerate and
mix in a complicated way. Fortunately, the operator mixing in
$\CN=4$ supersymmetric Yang-Mills (SYM) theory possesses rich hidden
symmetries that make the problem tractable. The one-loop planar
mixing matrix (dilatation operator) turns out to be a Hamiltonian of
an integrable quantum spin chain
\cite{Minahan:2002ve,Beisert:2003yb}. The spin-chain Hamiltonian is
a member of an infinite series of commuting charges and can be
diagonalized by powerful techniques from the Bethe ansatz. The
integrability in SYM extends to at least three loops
\cite{Beisert:2003tq,Klose:2003qc,Beisert:2003ys,Serban:2004jf} and
probably to higher orders of perturbation theory
\cite{Beisert:2004hm}. It is therefore natural to expect that the
dual string theory is integrable as well. Turning the argument
around, the AdS/CFT duality and the putative quantum integrability
of the AdS sigma-model would naturally explain the otherwise
miraculous integrability of the operator mixing in SYM
\cite{Dolan:2003uh}. The classical sigma-model on \ads\ is indeed
completely integrable \cite{Bena:2003wd,Polyakov:2004br}, but not
much is known about the quantum theory.

Even though integrable systems are incomparably simpler than
non-integrable, finding the spectrum of a quantum integrable model
is still a non-trivial task. To the best of our knowledge, the only
tool that possesses sufficient degree of universality is the Bethe
ansatz \cite{Bethe:1931hc,Thacker:1980ei,Faddeev:1996iy}. The
classic example of the model solvable by the Bethe ansatz is the
Heisenberg spin chain \cite{Bethe:1931hc,Faddeev:1996iy}. The Bethe
ansatz solution of this model and related spin systems was extremely
useful in comparing anomalous dimensions of local operators in SYM
\cite{Beisert:2003xu,Beisert:2003ea,Engquist:2003rn} to the energies
of classical string solitons in \ads\ \cite{Frolov:2003qc,
Arutyunov:2003uj,Arutyunov:2003za,Tseytlin:2003ii}. The energies
were found to agree with the scaling dimensions up to two loops in
many particular cases\footnote{The discrepancies found at three
loops for the BMN operators
\cite{Callan:2003xr,Callan:2004uv,Callan:2004ev} and for the
semiclassical string states \cite{Serban:2004jf}  can be attributed
to the weak/strong coupling nature of the AdS/CFT correspondence
\cite{Beisert:2004hm} that apparently manifests itself even in the
semiclassical regime \cite{Klebanov:2002mp}.}. Higher charges of
integrable hierarchies were also identified for particular solutions
\cite{Arutyunov:2003rg,Engquist:2004bx}. The relationship between
spin chains and the sigma-model was subsequently established quite
generally at the level of effective actions
\cite{Kruczenski:2003gt,Kruczenski:2004kw,Dimov:2004qv,
Hernandez:2004uw,Stefanski:2004cw,Ryzhov:2004nz,Kruczenski:2004cn},
equations of motion \cite{Mikhailov:2003gq}, or at the level of
Bethe ans\"atze \cite{Kazakov:2004qf}.

Although the Bethe ansatz is a purely quantum concept, it leaves
certain imprints in the classical dynamics. The classical solutions
of the sigma-model can be parameterized by an integral equation that
strikingly resembles the scaling limit of Bethe equations for the
spin chain \cite{Kazakov:2004qf}. In fact, the two equations become
equivalent at weak coupling. This observation lends strong support
to the idea that the quantum sigma-model is solvable by the Bethe
ansatz. The hypothetical exact Bethe equations for the sigma-model
should be discrete, as any quantum Bethe equations, and should
reduce to the integral equation derived in \cite{Kazakov:2004qf} in
the classical limit. A particular discretization of the classical
Bethe equation  of \cite{Kazakov:2004qf} was proposed in
\cite{Arutyunov:2004vx} and passed several non-trivial tests: the
equations of \cite{Arutyunov:2004vx} reproduce known quantum
corrections \cite{Callan:2003xr,Callan:2004uv,Callan:2004ev} to the
energies of BMN string states \cite{Berenstein:2002jq}
%, although
%apparently it is not a single possible discretization of the
%integral equations obeying this property.
and recover the $(g^2N)^{1/4}$ asymptotics \cite{Gubser:1998bc} of
anomalous dimensions at strong coupling. Interestingly, the string
Bethe equations have a spin chain interpretation
\cite{Beisert:2004jw}.
%\footnote{
%according to the comment of N.~Beisert mentioned in the footnote of
%\cite{Arutyunov:2004vx}}.
%The conjecture of \cite{Arutyunov:2004vx} is based on the
%previously proposed all-loop Bethe Ansatz equations on the SYM side
%derived in \cite{Beisert:2004hm} from the requirements
%of integrability and BMN scaling.
 The Bethe equations of
\cite{Arutyunov:2004vx} are asymptotic in the sense that they
require the 't~Hooft coupling and the R-charge to be large,
%The quest for the
%full quantum integrability of the AdS string thus still remains a
%challenge.
so deriving the full quantum Bethe ansatz for the sigma-model still
remains a challenge.

The classical Bethe equations  were obtained in
\cite{Kazakov:2004qf} for the simplest $SU(2)$ sector of the
 \ads\ sigma-model which is dual
to scalar operators of the form $\tr (Z^{J_1}W^{J_2}+{\rm
permutations})$, where $Z=\Phi_1+i\Phi_2$ and $W=\Phi_3+i\Phi_4$ are
two complex scalars of $\CN=4$ supermultiplet. This sector is closed
under renormalization because of the R-charge conservation
\cite{Beisert:2003jj,Beisert:2004ry}. On the string side, the
$SU(2)$ sector corresponds to strings confined in the $S^3\times
\R^1$ subspace of \ads. A string in this sector has two independent
angular momenta, which are identified with the R-charges $J_1$ and
$J_2$. The $\R^1$ direction corresponds to the global AdS time.

In this paper we shall analyze another closed sector with
non-compact $SL(2)$ symmetry group. The operators in this sector are
composed of an arbitrary number of light-cone covariant derivatives
acting on an arbitrary number of  scalar fields of one type:
\be \label{OPER}
{\cal O}
%= \tr \( \der^S Z^J \)
=\tr \der^{S_1}Z\ldots \der^{S_J}Z,\qquad S_1+\ldots+S_J=S.
%+{\rm permutations}
\ee
where $\der=D_0+ D_1$ and $D_\mu=\p_\mu+i[A_\mu,\ldots],\
\mu=0,1,2,3$. Large operators in this subsector are dual to
classical strings that propagate in $AdS_3\times S^1\subset
AdS_5\times S^5$. The string in $AdS_3$ has two
 independent charges, the Lorentz spin
$S$ and the dilatation charge $\Delta$. These charges label
representations of  $SO(2,2)$, the symmetry group of $AdS_3$. The
R-charge $J$ of the operator (\ref{OPER}) corresponds to the angular
momentum along $S^1$ and the dilatation charge  $\Delta$ of the
string maps to the scaling dimension of the dual operator.

%Many explicit string solutions with the $SL(2)$ symmetry are known.
Perhaps the simplest string solution with the $SL(2)$ symmetry is
the folded spinning string at the centre of AdS
\cite{Gubser:2002tv}. Its energy has the same parametric dependence
on the spin ($\Delta\propto\ln S$ at large $S$) as the perturbative
anomalous dimension of the operator (\ref{OPER}) with $J=2$, which
is now known up to three loops \cite{Kotikov:2004er}\footnote{The
three-loop result of \cite{Kotikov:2004er} relies  upon certain
structural assumptions and was extracted from the explicit
three-loop calculation in QCD \cite{Moch:2004pa}. It is consistent
with the predictions based on integrability \cite{Beisert:2003tq}
and with the direct calculations in \4N SYM \cite{Eden:2004ua}.}.
The coefficient of proportionality interpolates between power series
in $\lambda=g^2N$ at weak coupling and power series in
$1/\sqrt{\lambda}$ at strong coupling \cite{Frolov:2002av}. The
latter starts from the $O(\sqrt{\lambda})$ term
\cite{Gubser:2002tv}. The situation changes if in addition to
spinning in $AdS_5$ the string rotates in $S^5$ with the angular
momentum $J\sim S\gg 1$ \cite{Frolov:2002av}. The string energy is
then analytic in $\lambda/J^2$ and can be directly compared to the
anomalous dimension of an operator of the form (\ref{OPER})
\cite{Beisert:2003ea}. The one-loop results for the folded string
completely agree. The agreement was also established for the
pulsating strings (solutions found in \cite{Minahan:2002rc} and
further discussed in \cite{Khan:2003sm})\cite{Smedback:1998yn}. The
relationship between effective actions for strings and spins in the
$SL(2)$ sector was derived in \cite{Stefanski:2004cw} and was
further studied in \cite{Bellucci:2004qr,Ryang:2004pu}. In this
paper we shall focus on the relationship between integrable
structures.

There is an important difference between the $SU(2)$ and $SL(2)$
sectors. In the former case the dilatation charge is the energy of
the string and is decoupled from the rest of the dynamics. In the
latter case the dilatation generator is a part of the $SO(2,2)$
isometry group and has non-trivial commutation relations with other
generators. In this sense the dilatation generator is not much
different from other global charges that together form a closed
symmetry algebra. This line of thought has proven extremely useful
for constructing the dilatation operator on the field-theory side of
AdS/CFT \cite{Beisert:2003ys,Beisert:2004ry}.

In section 2 we overview the Bethe ansatz solution for the one loop
dilatation operator in the $SL(2)$ sector of the $N=4$ SYM theory.
Then we derive the  classical limit for long operators. The Bethe
equations reduce to a Riemann-Hilbert problem in this limit.

In section 3 we describe the so called finite gap solutions of the
classical string rotating on the $AdS_3\times S_1$ space based on
the integrability. The problem  is again  reduced to a solvable
Riemann-Hilbert problem for the quasimomentum defined on a two-sheet
Riemann surface. The comparison of two Riemann-Hilbert problems in
the week coupling region shows the complete one-loop equivalence of
the gauge theory and the sigma model.

Sdection 4 contains the general solution of the one loop Bethe
equations in the classical limit. The general case is exemplified by
the rational solution, which is dual to the circular string
\cite{Arutyunov:2003za}. In section 5 the complete solution is
constructed for the sigma model. The rational case is treated in
some detail. Section 6 is devoted to the discussion.
%The comparison of two rational solutions shows again the
%agreement at one loop.

%%%%%%%%%%%%%%%%%%%%%%%%%%%%%%%%%%%%%%%%
\sectiono{Bethe Ansatz}\label{bethesec}

The operators (\ref{OPER}) with the same $J$ and $S$ are degenerate
at tree level. This degeneracy is lifted by quantum corrections. The
conformal operators with definite scaling dimensions are linear
combinations of basic operators (\ref{OPER}) with coefficients that
can be computed order by order in perturbation theory. At each
order, the conformal operators are eigenvectors of the mixing
matrix, whose eigenvalues are the corresponding anomalous
dimensions. The size of the mixing matrix rapidly grows with $S$ and
$J$, but the problem significantly simplifies at large $N$ when the
mixing matrix takes the form of an $sl(2)$ spin chain with $J$
sites. The operators are the states of the spin chain. Each entry
$\der^{S_l}Z$ in an operator  corresponds to a site of a
one-dimensional lattice. The sites are cyclically ordered because of
the overall trace. $Z$ without derivatives corresponds to an empty
site and $\der^{S_l}Z$ corresponds to a site in the $S_l$-th excited
state. The excitations are naturally classified according to the
infinite-dimensional spin $s=-1/2$ representation of $sl(2)$. The
mixing matrix acts pairwise on the nearest-neighbor sites of the
lattice and turns out to coincide with the Hamiltonian of the
integrable spin $s=-1/2$ XXX spin chain
\cite{Beisert:2003jj,Beisert:2003yb} which is similar to the spin
$s=-1$
\cite{Braun:1998id,Braun:1999te,Belitsky:2004cz,Belitsky:2004sc} and
$s=-3/2$ \cite{Belitsky:1999bf,Belitsky:2004cz,Belitsky:2004sc}
chains that describe anomalous dimensions of quasipartonic operators
in QCD. The spin chain is solvable by the Bethe ansatz, and the
spectrum of anomalous dimensions can be found by solving a set of
algebraic equations: \be\label{BETHE}
\left(\frac{u_j-i/2}{u_j+i/2}\right)^J=\prod_{k\neq j}
\frac{u_j-u_k+i}{u_j-u_k-i}\, \ee The roots $u_j$, $j=1,\ldots, S$
are distinct real numbers. The solutions of Bethe equations that
correspond to eigenstates of the mixing matrix satisfy an additional
constraint \be\label{MOME} \prod_j\frac{u_j-i/2}{u_j+i/2}=1. \ee
This condition takes into account the cyclicity of the trace in
(\ref{OPER}). Bethe states that satisfy this condition have zero
total momentum and are invariant under cyclic permutations of the
elementary fields. For a given solution of the Bethe equations,
 the anomalous
dimension is determined by \be
\Delta=S+J+\frac{\lambda}{8\pi^2}\sum_j\frac{1}{u_j^2+1/4}
+O(\lambda^2).\ee More details about the Bethe ansatz and its
relationship to the anomalous dimensions of $sl(2)$ operators can be
found in \cite{Beisert:2003yb,Beisert:2003jj}.

We are interested in the scaling limit $S\rightarrow\infty$,
$J\rightarrow\infty$ with the ratio $S/J$ held fixed. This scaling
limit was discussed for the $SU(2)$ sector in
\cite{s1,s2,Beisert:2003xu,Kazakov:2004qf}. The $SL(2)$ case can be
understood as an analytic continuation in the spin
\cite{Beisert:2003ea}, though there are some differences in the
reality conditions for Bethe roots. The Bethe roots scale with $J$
as $u_j\sim J$. Equating the phases of both sides of (\ref{BETHE})
and expanding in $1/u_j$ we get
\be\label{BETHEs} \sum_{k\neq j}\frac{2}{u_j-u_k}=2\pi
n_j-\frac{J}{u_j}\,. \ee
The mode numbers $n_j$ arise because one can choose different
branches of the logarithm for different Bethe roots. We shall assume
that a macroscopic ($\sim J$) number of Bethe roots have equal mode
numbers. The distribution of Bethe roots then can be characterized
by a continuous density
\be\label{DENS}
\rho(x)=\sum_{j=1}^S\delta\left(x-\frac{u_j}{J}\right). \ee
 The
density has a support on a set of disconnected intervals
$C_1,\ldots,C_K$ of the real axis. The interval $C_i$ is filled by
roots with the mode number $n_i$ and is centered around $x=1/2\pi
n_i$. We can also define the resolvent
\be\label{RES}  G(x)=\sum_{j=1}^S\frac{1}{Jx-u_j} =\int
dy\,\frac{\rho(y)}{x-y}, \ee
which is an analytic function of $x$ on the complex plane with cuts
along the intervals $C_i$. The density, according to the definition
\rf{DENS}, is normalized as
\be\label{NORM} \left. xG(x)\right|_{x=\infty}=\int
dx\,\rho(x)=\frac{S}{J}. \ee

The scaling limit of Bethe equations translates into an integral
equation for the density:
\be\label{Bint} 2\pint dy\,\frac{\rho(y)}{x-y}=2\pi
n_i-\frac{1}{x}\qquad {\rm for~}x\in C_i, \ee
or, in terms of the resolvant, \be\label{Bint1} G(x+i0)+G(x-i0)=2\pi
n_i-\frac{1}{x}\qquad {\rm for~}x\in C_i. \ee It is also useful to
introduce the quasi-momentum
\be\label{pG} p(x)=G(x)+{1\over 2x}\,  \ee
which satisfies
\be \label{PPP} p(x+i0)+p(x-i0)=2\pi  n_i,\qquad x\in  C_i \ee
 The momentum condition  (\ref{MOME})  constraints the first
moment of the density:
\be \label{momint} \int dx\,\frac{\rho(x)}{x}=-2\pi m, \ee
 where $m$ is an arbitrary
integer. The second moment determines the anomalous dimension:
\be\label{engint} \Delta-S-J=\frac{\lambda}{8\pi^2J}\int
dx\,\frac{\rho(x)}{x^2}\,. \ee
 The general solution of the integral equation (\ref{Bint}) is
derived in sec.~\ref{ss:kri}. In the next section we shall derive
equations analogous to (\ref{Bint}), (\ref{NORM}), (\ref{momint})
and (\ref{engint}) in the classical sigma-model.

%%%%%%%%%%%%%%%%%%%%%%%%%%%%%%%%%%%%%%%%%%%%%%%%%%%%%%%%%%%%%%%%%%%
\sectiono{ Classical Strings on $AdS_3\times S^1$ }\label{ss:classic}
%%%%%%%%%%%%%%%%%%%%%%%%%%%%%%%%%%%%%%%%%%%%%%%%%%%%%%%%%%%%%%%%%%%

%%%%%%%%%%%%%%%%%%%%%%%%%%%%%%%%%%%%%%%%%%%%%%%%%%%%%%%%%%%%%%%%%%%%%%
\subsection{The model}
\label{ss:sigmam}
%%%%%%%%%%%%%%%%%%%%%%%%%%%%%%%%%%%%%%%%%%%%%%%%%%%%%%%%%%%%%%%%%%%%%%

As in the discussion of the SYM operators we shall focus on a
particular reduction of the full $AdS_5\times S^5$ sigma-model by
considering strings that move in $AdS_3\times\R^1$.
%We will consider classical solution of the sigma-model
%which describe a string moving on $AdS_3\times%\mathbb
%S_1$.
The symmetry algebra of the sigma-model on
 $AdS_3$ is $so(2,2)\sim sl(2)\times sl(2)$. The
$AdS_3$  space is the group manifold of $SL(2,\R)$ and the two
$sl(2)$ symmetries act as the left and right group multiplications.
We should mention that the $AdS_3$ sigma-model with the WZW term is
rather well understood \cite{Maldacena:2000hw} but has quite
different properties, even at the classical level. The background
NS-NS flux of the WZW model couples directly to the classical string
world-sheet, unlike the R-R flux of the \ads \ background that is
only important in quantum theory.

The string action in the conformal gauge is\footnote{The signature
of the world-sheet metric is $(+-)$. The effective string tension is
related to the 't~Hooft parameter according to the AdS/CFT
correspondence \cite{Maldacena:1998re}.}
\be \label{SIGMA}
S_{\sigma}% = S+S_0 =
={\sqrt{\lambda}\over 4\pi }\int_0^{2\pi} d\sigma\, \int d\tau\,
\left[ -\d_aX_i\d^aX^i +\left(\d_a \phi\right)^2\right], \ee
 where
$\phi$ is the angle on a big circle of $S_5$ and $X_i$, $i=-1,0,1,2$
are the $AdS_3$ embedding  coordinates. They parameterize a
hyperboloid in the four-dimensional space with the signature
$(++--)$:
\be\label{sphere}
X_iX^i=X_{-1}^2+X_0^2-X_1^2-X_2^2=X_+X_-+Y_+Y_-=1,
\ee
where we introduced $X_\pm=X_{-1}\pm X_1$, $Y_\pm= X_0\pm X_2$. All
other world-sheet coordinates on $AdS_5$ and $S^5$ are set to
constant values. Classically, this is a consistent reduction. The
$SO(2,2)$ symmetry is manifest in the above parametrization.

The equations of
motion that follow from \rf{SIGMA} should be supplemented by
Virasoro constraints:
\be \d_\pm X_i\d_\pm X^i=(\d_\pm \phi)^2.
\ee
where $\sigma_\pm=(\tau\pm\sigma)/2$, $\d_\pm=\d_\tau\pm\d_\sigma$.
We can always choose the gauge
$$\phi={J\over\sqrt{\lambda}}\,\tau+m\sigma,$$
where $m$ is the winding number\footnote{The circular string
solutions with the non-zero winding were constructed in
\cite{Arutyunov:2003za}. The appearance of the winding around the
decoupled $S^1$ factor is a novel feature of the $AdS_3\times S^1$
background compared to the $S^3\times \R^1$ case. We would like to
thank A.~Tseytlin for the discussion of this point.},
 then
\be\label{VIRA} \d_\pm X_i\d_\pm
X^i=\left(\frac{J}{\sqrt{\lambda}}\pm m\right)^2. \ee The angular
momentum on $S^5$,
\be\label{MOM} J=\frac{\sqrt{\lambda}}{2\pi}\int_0^{2\pi} d\sigma\,
\d_0 \phi, \ee
should be identified with the number of the $Z$ fields in the
operator \rf{OPER}.

A point in $AdS_3$ defines a group element of $SL(2,\R)$:
%\footnote{we use here the notations of \cite{Maldacena:2000hw}}
%
\be \label{SODD}
 g = X_{-1}+{\bf X}\cdot
 \sm \equiv \left(
\begin{array}{cc}
  X_{-1}+X_1 & X_0-X_2 \\
 -X_0-X_2 & X_{-1}-X_1
\end{array}
\right) \equiv \left(
\begin{array}{cc}
  X_+ & Y_- \\
  -Y_+ &  X_-
\end{array}\right) \in SL(2,\R),
\ee
where ${\bf X}=(X_0,X_1,X_2)$
%we choose the basis of the Pauli
%matrices
 and $\sm =(i\s_2,\s_3,-\s_1)$.
% Let us parametrize the $SO(2,2)$ group element by the
Another useful parametrization of an $SL(2,\R)$ group element is
%Euler angles $u,v$ and the exponential coordinate in the non-compact
%direction $\rho$ as follows:
%
\be \label{ANGLES} g=e^{iu\s_2}e^{\rho\s_3}e^{iv\s_2}=\left(
\begin{array}{cc}
 \cos t  \cosh\rho+\cos\psi\sinh\rho &
 \sin t \cosh\rho-\sin\psi\sinh\rho \\
 -\sin t \cosh\rho-\sin\psi\sinh\rho &
 \cos t  \cosh\rho-\cos\psi\sinh\rho
\end{array}
\right) \ee
where $t$ is the global AdS time, $\rho$ is the radial variable and
$\psi$ is an angle. $u$ and $v$ are the light-cone coordinates:
\be\label{LCQ}   u=\hf(t+\psi),\qquad v=\hf(t-\psi). \ee
%
%where $t$ plays the role of the  renorm-group "time" conjugated to
%the dilaton Hamiltonian.
%Note that $det g=1$ and $g^{-1}\s_3g=\s_3$.
The differential on the group manifold has the following form:
\be \label{DIFF} g^{-1}dg = \left(
\begin{array}{cc}
 X_- dX_+ - Y_+dY_+ &  X_- dY_- + Y_+dX_- \\
  -Y_- dX_+ - X_+dY_+  & -Y_- dY_- + X_+dX_-
\end{array}\right) \in sl(2,\R).
\ee
The invariant metric then is
\be\label{ADS}  ds^2=-\frac{1}{2}\tr(g^{-1}dg)^2
=dX_{-1}^2+dX_0^2-dX_1^2-dX_2^2 =\cosh^2\rho\ d t^2-d\rho^2
-\sinh^2\rho\ d\psi^2 \ee
%
%The metric in the r.h.s.  covers globally the whole $AdS_3$ space.
%The angle $\psi$ corresponds to a compact angle of the $AdS_3$, and

The time coordinate is an angular variable in the parameterization
(\ref{ANGLES}). As a consequence, $t(\sigma ,\tau )$ is not
necessarily periodic in $\sigma $ even if $g(\sigma, \tau )$ is
periodic. This makes boundary conditions a non-trivial issue. Just
requiring that $g(\sigma+2\pi , \tau )=g(\sigma, \tau )$ is not
sufficient because this condition allows the string to wind around
the time direction\footnote{We would like to thank A.~Tseytlin and
S.~Frolov for drawing our attention to this fact.}. We will return
to the issue of the time-like windings later.

%The isometries of $AdS_3$ are generated by left and right group
%multiplication.
Let us now figure out which global charges in the SYM correspond to
Noether charges of the left and right group multiplications in
$SL(2,\R)$.
 The boundary of $AdS_3$ is
located at $\rho\to\infty$.  Asymptotically, the metric takes the
form
\be\label{ADS1}  ds^2 ={e^{2\rho}\over r^2}\( dr^2 -
r^2d\psi^2\),\qquad  t=\log r,  \ee
and is conformal to the two-dimensional Minkowski metric. The
rescalings $r\to \Lambda r$ or, in the original variables, $t\to
 t+\eps$ act as dilatations
on the boundary. The associated conserved charge should be
identified with the scaling dimension $\D$ of the operator
(\ref{OPER}). The $U(1)$ rotations $\psi\to\psi+\eps'$ correspond to
boosts in  $x^1$ direction, under which $D_+$ in (\ref{OPER})
transforms as $D_+\to \e^{-i\eps'}D_+$. The $U(1)$ charge in the
sigma-model thus corresponds to the spin $S$ of the SYM operator.

In the representation (\ref{ANGLES}), the scaling transformations
 correspond to
simultaneous left and right multiplication by
$\e^{i\eps\sigma_2/2}$. The boosts are generated by the left
multiplication by $\e^{i\eps'\sigma_2/2}$ and the right
multiplication by $\e^{-i\eps'\sigma_2/2}$. The Noether currents of
left/right group multiplications are
\be\label{CURR} j_a=g^{-1}\d_a g={1\over 2}\,j^A_a s^A,\qquad
 l_a=g^{-1}j_a g=\p_a gg^{-1}={1\over
2}\,l^A_a s^A,\qquad a=0,1=\tau,\sigma. \ee
%
%from these identifications
Therefore,
\be\label{CHRGES} \Delta+S =\frac{\sqrt{\lambda}}{4\pi}\int_0^{2\pi}
d\sigma\, j^0_0,\qquad  \Delta-S
=\frac{\sqrt{\lambda}}{4\pi}\int_0^{2\pi} d\sigma\, l^0_0.
 \ee
Finally, the Virasoro constraints \rf{VIRA} become
\be \label{VIRAS} \ha \tr j_\pm^2=-\left(\frac{J}{\sqrt{\lambda}}\pm
m\right)^2 , \ee
where $j_\pm = g^{-1}\d_\pm g$.

%%%%%%%%%%%%%%%%%%%%%%%%%%%%%%%%%%%%%%%%%%%%%%%%%%%%%%%%%%%%%%%%%%%
\subsection{Equations of motion and integrability}
\label{ss:chiral}
%%%%%%%%%%%%%%%%%%%%%%%%%%%%%%%%%%%%%%%%%%%%%%%%%%%%%%%%%%%%%%%%%%%

We shall analyze the classical solutions of the $\s$-model on
$AdS_3\times S^1$ along the same lines as solutions of the $SU(2)$
sigma model were analyzed in \cite{Kazakov:2004qf}. According to
(\ref{ADS}), we can write the action \rf{SIGMA} in the form
\be \label{cf} S_{\sigma }= \frac{\sqrt{\lambda}}{4\pi} \int d\sigma
d\tau\, \left[\frac{1}{2}\,\tr j_a^2+(\d_a \phi)^2\right]. \ee
The equation of motion for the $S^1$ coordinate is just the Laplace
equation
$$\d_+\d_- \phi=0$$
and is solved by $\phi=J\tau/\sqrt{\lambda}+m\sigma$.

 The equations of motion for the $sl(2)$ currents can be
written as follows
\be\label{sm/eq}      \d_+j_-+\d_-j_+=0, \qquad
\d_+j_--\d_-j_++[j_+,j_-]=0. \ee
where the last equation is a consequence of the definition
\rf{CURR}.
%The current $j_a$ is a pure gauge by definition and is
%conserved on shell.
The equations of motion  can be reformulated as the flatness
condition \cite{Zakharov:pp} for a one-parametric family of currents
$J(x)$: \be J_\pm(x) = \frac{j_\pm}{1\mp x}\,. \ee If \rf{sm/eq} are
satisfied, then \be\label{zerocur} \d_+J_--\d_-J_++[J_+,J_-]=0. \ee
The converse is also true. If the connection $J_a(x)$ is flat for
any $x$, the current $j_a$ solves the equations of motion
\rf{sm/eq}.

The zero-curvature representation effectively linearizes the
problem. Instead of analyzing the equations of motion, which are
non-linear, we can study the auxiliary linear problem:
\be\label{laxL} {\cal L}\Psi \equiv \left(\d_\sigma +
\ha\left({j_+\over 1-x} - {j_-\over 1+x}\right)\right)\Psi =0, \ee
\be\label{laxM} {\cal M}\Psi \equiv \left(\d_\tau +
\ha\left({j_+\over 1-x} + {j_-\over 1+x}\right)\right)\Psi =0, \ee
for which \rf{zerocur} is the consistency condition.
%The solution to the linear problem   has essential singularities at
%$x \to \pm 1$, where near the singularity it has a solution similar
%to \rf{baf}, and
%
%The solutions of the equations of motion \rf{cf} can be
%reconstructed  from the solution of the auxiliary linear problem:
%
%\be\label{SOLCF}    j_\pm=(x\mp 1)\(\Psi_\tau \Psi^{-1}\mp \Psi_\s
%\Psi^{-1}\). \ee
%

The  solution of \rf{laxL} with the initial condition
$\Psi(\tau,0)=1$
 defines the monodromy matrix:
\be \Omega(x)=P\exp\int_0^{2\pi} d\sigma\, \ha\left( {j_-\over
1+x}-{j_+\over 1-x}\right), \ee
%
% Since $\Omega\in SL(2)$, it has  real  eigenvalues for for real $x$,
%   $(e^{\pm ip(x)}$, thus defining  the quasi-momentum $p(x)$
and the quasi-momentum $p(x)$:
\be\label{LAMDAOMEGA} \tr\Omega(x)=2\cos p(x). \ee
Since the trace of the holonomy of a flat connection does not depend
on the contour of integration, the quasi-momentum does not depend on
$\tau$, in other words, $p(x)$ is conserved. The quasi-momentum
$p(x)$ depends on a parameter and thus generates an infinite set of
integrals of motion, for instance by Taylor expansion in $x$.

The complete linear problem (\ref{laxL}), (\ref{laxM}) and hence the
solution of the original non-linear equations can be reconstructed
from the quasi-momentum provided that it satisfies certain
analyticity conditions. The procedure, sometimes called the
inverse-scattering transformation, is described in detail in
\cite{book_of_soliton}. We will not use the full machinery of this
method. It will be sufficient for our purposes to derive the
analyticity constraints on the quasi-momentum as a function of the
spectral parameter.

%%%%%%%%%%%%%%%%%%%%%%%%%%%%%%%%%%%%%%%%%%%%%%%%%%%%%%%%%%%%%%%%%%
\subsection{Analytic properties of the quasi-momentum }
\label{ss:analyt}
%%%%%%%%%%%%%%%%%%%%%%%%%%%%%%%%%%%%%%%%%%%%%%%%%%%%%%%%%%%%%%

%As in \cite{Kazakov:2004qf}\footnote{See there the details of the
%construction}, we will follow the procedure to obtain the finite-gap
%solutions to sigma models proposed in \cite{Krisig}. It uses the
%functions on the double cover of the spectral curve. The details of
%this  construction and a particular (rational) solution for the
%non-compact sigma-model considered in this paper are given in
%Appendix~\ref{ss:kri}.

%Note $\Psi(x;\tau,\sigma)$ is essentially determined by the analytic
%properties of the quasi-momentum $p(x)$.
%
%Let us now establish the analytical properties of the quasi-momentum
%by analyzing the auxiliary linear problem.
Our exposition closely follows \cite{book_of_soliton} and largely
repeats the analysis of the $SU(2)$ sigma-model
\cite{Kazakov:2004qf}. There are however some important
modifications due to the non-compactness of the target space. The
auxiliary problem
\be\label{laxL1} %\L\Psi =
\left[\d_\sigma + \ha\left({j_+\over 1-x} - {j_-\over
1+x}\right)\right]\psi =0 \ee resembles one-dimensional Dirac
equations (now $\psi$ is a column vector as opposed to (\ref{laxL}),
where $\Psi$ was a two-by-two matrix). It has
 two linearly independent solutions which
can be chosen quasi-periodic.
%This is a common place, but it is
It is useful to see how quasi-periodicity is related to the
monodromy matrix. If the initial conditions are its eigenvalues
 \be
\Omega(x)\psi_\pm(x;0)=\e^{\pm i p(x)}\psi_\pm(x;0), \ee
 the
solution $\psi_\pm(x;\sigma)=\Psi(x;\sigma)\psi_\pm(x;0)$ will
satisfy $\psi_\pm(x;\sigma+2\pi)=\e^{\pm  ip(x)}\psi_\pm(x;\sigma)$
because $\Psi(x;\sigma+2\pi)=\Psi(x;\sigma)\Omega(x)$.

Since the monodromy matrix $\Omega(x)\in SL(2,\R)$, $\cos p(x)$ is
real for real $x$, but the quasi-momentum itself is not necessarily
real. The condition for that is $\tr\Omega(x)\leq 2$. Then the
quasi-periodic solutions are delta-function normalizable. This
corresponds to allowed zones of the one-dimensional Dirac equation
(\ref{laxL1}). In  the forbidden zones
 $\tr\Omega(x)\geq 2$, the quasi-momentum is imaginary and the wave
 functions grow exponentially at infinity.
The number of forbidden zones is in general infinite, but there is a
representative set of solutions (finite-gap solutions) for which
this number is finite.

The quasi-momentum $p(x)$ can be analytically continued to complex
values of $x$. Its only singularities are at zone boundaries and at
$x=\pm 1$, where the potential in (\ref{laxL1}) is singular.
Therefore $p(x)$ is a meromorphic function on the complex plane with
cuts along the forbidden zones. Let us explain why zone boundaries
are branch points. The monodromy matrix generically has two distinct
eigenvalues, but at zone boundaries it degenerates into the Jordan
cell and has only one eigenvector with the eigenvalue $1$ or $-1$.
The quasi-momentum becomes an integer multiple of $\pi$ such that
the single quasi-periodic solution of \rf{laxL} is either periodic
or anti-periodic\footnote{The Dirac equation may also have two
linearly independent periodic or anti-periodic solutions at isolated
points in the $x$ plane. Such double points should not be confused
with zone boundaries, where the Dirac equation has only one
(anti)periodic solution. If $x_0$ is a double point, then
$p(x_0)=\pi n,\ n=0,\pm 1,\pm 2,\ldots$ and $dp(x_0)=0$. The double
points can be viewed as forbidden zones shrunk to zero size.}. Two
linearly independent solutions of (\ref{laxL1}), $\psi_+(x,\sigma)$
and $\psi_-(x,\sigma)$, collapse into one degenerate solution at
zone boundaries and are analytic functions of $x$ elsewhere (except
for at $x=\pm 1$). Thus $\psi_+(x,\sigma)$ and $\psi_-(x,\sigma)$
behave precisely as two branches of a single meromorphic function on
a double cover of the complex $x$ plane. The two eigenvalues of the
monodromy matrix, $\e^{\pm i p(x)}$, are also branches of a single
meromorphic function on the hyperelliptic surface the two sheets of
which are glued together along the forbidden zones.  Another way to
see that the quasi-momentum is naturally defined on the Riemann
surface is to notice that (\ref{LAMDAOMEGA}) is a quadratic equation
for $\e^{ip(x)}$. The trace of the monodromy matrix is an entire
function of $x$, but its eigenvalues have square root singularities
when the discriminant of the equation \rf{LAMDAOMEGA} turns to zero,
and this happens precisely at zone boundaries when $\tr\Omega=2$.

To summarize, the eigenvalues of the monodromy matrix $\e^{\pm
ip(x)}$ are branches of a single meromorphic function on the
hyperelliptic Riemann surface. A particular branch $p(x)$ is
analytic on the complex plane with cuts along the forbidden zones.
We shall identify these cuts with the intervals on which Bethe roots
of the spin chain condense.
% The quasi-momentum is real (modulo $2\pi i$)
%on a set of disjoint one-dimensional supports
%(some segments in the complex plane of $x$). These segments lie on
%the real axis and  correspond to allowed zones of the spectral
%problem.  The allowed zones are singled out by the condition that
%${\rm Im}\,\left({\rm Tr}\,\Omega\right)=0$ and $|{\rm
%Tr}\,\Omega|>2$. One can also define forbidden zones as loci where
%${\rm Im}\,\left({\rm Tr}\,\Omega\right)=0$ and $|{\rm
%Tr}\,\Omega|<2$. In general the number of allowed and forbidden
%zones is infinite. We will only discuss finite gap, or
%algebro-geometric solutions for which this number is finite. They
%are governed by the geometry of a complex curve of finite genus and
%directly correspond to solutions of the Bethe equations with a
%finite number of cuts. All other solutions can be probably obtained
%as limiting cases of the finite gap solutions, when the number of
%the forbidden zones increases to infinity.

%%%%%%%%%%%%%%%%%%%%%%%%%%%%%%%%%%%%%%%%%%%%%
\subsection{Finite gap solution and asymptotic conditions}
%%%%%%%%%%%%%%%%%%%%%%%%%%%%%%%%%%%%%%%%%%%%

Consider now the behavior of the quasi-momentum near one of the
forbidden zones. The values of $\e^{ip(x)}$ on the two sides of the
cut, $\e^{ip(x+i0)}$ and $\e^{ip(x-i0)}$, are two independent
solutions of \rf{LAMDAOMEGA}. Since $\Omega(x)$ is unimodular,
$\e^{ip(x+i0)}\e^{ip(x-i0)}=1$, and the quasi-momentum satisfies the
equation equivalent to \rf{PPP}:
\be \label{RHPR} p(x+i0)+p(x-i0)=2\pi  n_k,\qquad x\in C_k, \ee
which holds on each of the forbidden zones.
%The quasi-momentum can be
%related to the "resolvent" $G_s(x)=\int{dy\ \rho_s(y)\over x-y}$:
%
%\be\label{pG} p(x)=-i\(G_s(x)+{x\over 2(x^2-1)}\)   \ee
%
%having no singularities but the cuts along the forbidden zones on
%the physical sheet. Then we can write instead of \rf{PPP} the
%integral equation on the the density $\rho_s(x)$ similar to
%\rf{Bint}.
The integer $n_k-n_{k-1}-1$ is the number of (anti)-periodic
solutions within the $k$-th allowed zone, that is, the number of
 the double points between $C_{k-1}$ and $C_k$.

% Since the derivative of the quasi-momentum
% $p(x)$ is single-valued outside of the cuts, all ${\bf
%A}$-periods are zero:
%
%\be\label{AKPER} \oint_{A_k}dp=0,\qquad  x\in {\bf C}_k\ee
%

%There are no "condensate" cuts corresponding to the "strings" of
%equally distanced rapidities. As was remarked in
%\cite{Kazakov:2004qf}, it is always possible to choose cuts in such
%a way that condensates "evaporate'' and all ${\bf A}$-periods of
%$dp$ identically turn to zero. Such a choice of  cuts is unique and
%corresponds precisely to cutting the $x$-plane along the forbidden
%zones.

 The auxiliary linear problem \rf{laxL}, \rf{laxM}  becomes singular
 at $x=\pm 1$ and the quasi-momentum develops a pole there.
 The standard asymptotic  analysis yields
\be\label{pres1} p(x)=\pi\, \frac{\frac{ J}{\sqrt{\lambda}}\mp
m}{x\pm 1}+\ldots \qquad (x\rightarrow \mp 1). \ee
 It can be justified by
dropping the non-singular pole term in  \rf{laxL1}, writing the
Schr\"odinger type equation for one of the two components of $\psi$
and solving it in the WKB approximation. The asymptotic analysis
determines $p(x)$ only up to a sign. Fixing the sign ambiguity, as
in (\ref{pres1}), excludes a part of  solutions, for example
pulsating strings of
\cite{Minahan:2002rc,Khan:2003sm,Smedback:1998yn}. This point is
discussed in more detail in sec.~5.3 of \cite{Kazakov:2004qf}.

 To express the charges in terms of the
spectral data, we expand the quasi-momentum at zero and at infinity.
At infinity, ${\cal L}=\partial_\sigma-j_0/x+\ldots$, and
\be \Tr\Omega=2+\frac{1}{2x^2}\int_0^{2\pi}d\sigma_1 d\sigma_2\, \Tr
j_0(\sigma_1) j_0(\sigma_2)+\ldots
%=2-\frac{4\pi^2 Q_R^2}{\lambda x^2}+\ldots
=2-\frac{4\pi^2 (\D+S)^2}{\lambda x^2}+\ldots \,. \ee
Here we assume that the classical solutions correspond to
highest-weight states and use (\ref{CHRGES}).
%For highest weights in large
%representations, $|Q_{L,R}|^2$ can be replaced with
%From \rf{CHRGES} we have $(Q_{L}^3)^2=(\D-S)^2$ and
%$(Q_{R}^3)^2=(\D+S)^2$.
Thus
\be\label{XINF}   p(x)=\frac{2\pi(\D+S)}{\sqrt{\lambda}\,
x}+\ldots\qquad (x\rightarrow\infty). \ee

At $x\to 0$, ${\cal L}=\partial_\sigma+j_1-x j_0+\ldots$, which can
be written as ${\cal L}=g^{-1}(\partial_\sigma-x l_0+\ldots)g$.
Then,
$$
\Omega(x)=g^{-1}(2\pi)P
\exp\left(x\int_0^{2\pi}d\sigma\,l_0+\ldots\right)g(0).
$$
Because $g(\sigma)$ is periodic, $g(2\pi)=g(0)$, and thus
$\Omega(0)=1$. As we discussed in sec.~\ref{ss:sigmam}, the
periodicity of $g(\sigma)$ does not guarantee the periodicity of the
AdS time coordinate. The time coordinate is an angular variable in
the $SL(2,\R)$ parameterization of $AdS_3$ and we need to eliminate
the unphysical time-like windings by hand. It is easy to see that
the integer $p(0)/2\pi $ is precisely the winding number around the
time direction: in the simplest case of the string in the middle of
AdS ($\rho =0$), $j_1=\partial_\sigma  t$ and $p(0)=t(2\pi )-t(0)$.
We thus keep only the solutions with
 $p(0)=0$.
Expanding the quasi-momentum further, we get
\be \Tr\Omega=2+\frac{x^2}{2}\int_0^{2\pi}d\sigma_1 d\sigma_2\, \Tr
l_0(\sigma_1) l_0(\sigma_2)+\ldots
%=2+\frac{4\pi^2Q_L^2}{\lambda}\,x^2 +\ldots
=2-\frac{4\pi^2(\D-S)^2}{\lambda }\,x^2 +\ldots\,. \ee
Hence,
\be\label{SMALLX} p(x)=-\frac{2\pi (\D-S)}{\sqrt{\lambda} }\,
x+\ldots,\qquad (x\rightarrow 0). \ee

The quasi-momentum is a meromorphic function on the complex plane
with cuts and has two poles at $x=\pm 1$.
 Subtracting the
singularities at $x\to\pm 1$, we get the function
\be G(x)=p(x)-\pi\left(\frac{\frac{
J}{\sqrt{\lambda}}+m}{x-1}+\frac{\frac{
J}{\sqrt{\lambda}}-m}{x+1}\right), \ee
which is analytic everywhere on the physical sheet.
% Because
%the poles at $x=\pm 1$ cancel, the resolvent is an analytic function
%on the physical sheet and, as in \rf{RES}, it can be represented as
%an integral of a positive density
As such, it admits a spectral representation where the spectral
density is the discontinuity of $G(x)$ across the cuts: $\rho(x)=\Im
G(x)/\pi$. The standard analyticity arguments yield
\be\label{spectrepr} G(x)=\int d\xi\,\frac{\rho(\xi)}{x-\xi}\,. \ee
%
% The proof of this spectral
%representation is the standard argument based on analyticity of
%$G(x)$. We define $2i\pi\rho(x)=G(x+i0)-G(x-i0)$. Then the
%right-hand side can be represented by a contour integral with the
%contour surrounding  all the cuts. The only singularity of the
%integrand on the outside of the contour is a pole at  $y=x$ with
%residue $\rho(x)$. Shrinking the contour, we get \rf{spectrepr}.

  The asymptotic behavior of the resolvent at $x\to\infty$
is determined by \rf{XINF}: $G(x)\sim 2\pi[(\D+S-J)/\sqrt{\lambda
}]/x$, and
 translates into the normalization condition for the density:
\be\label{nn} \int dx\,\rho(x) =\frac{2\pi}{\sqrt{\lambda}}(\D+S-J).
\ee
The asymptotics at $x\to 0$ follows from \rf{SMALLX} and yields two
other conditions:
\be\label{anomalous-dim}
 \int dx\,\frac{\rho(x)}{x^2}=\frac{2\pi}{\sqrt{\lambda}}
(\Delta-S-J) \ee
and
 \be\label{cbethemom}  \int dx\,\frac{\rho(x)}{x}=-2\pi m. \ee
The spectral representation \rf{spectrepr} and the equation
\rf{RHPR} imply that the density satisfies a singular integral
equation:
\be\label{cbethe}  2\pint
dy\,\frac{\rho(y)}{x-y}=-2\pi\left(\frac{\frac{
J}{\sqrt{\lambda}}+m}{x-1} +\frac{\frac{
J}{\sqrt{\lambda}}-m}{x+1}\right)+2\pi n_k, \qquad x\in {\bf C}_k.
\ee
We obtained a  Riemann-Hilbert problem similar to the one appeared
in \cite{Kazakov:2004qf} for the $SU(2)$ sector.
%The only difference
%with the similar equation is the opposite sign of the pole terms in
%\rf{cbethe}.
In fact, if we set the winding number to zero, the equations for the
$SU(2)$ sectors can be obtained from the equations for $SL(2,\R)$ by
an analytic continuation that first appeared in the analysis of
particular solutions \cite{Beisert:2003ea}: $J\rightarrow\Delta$,
$S\rightarrow J_2$, $\Delta\rightarrow -J_1$\footnote{$J_2=J$,
$J_1=L-J$ in the notations of \cite{Kazakov:2004qf}.}. This duality
is a consequence of the fact that the AdS space can be obtained from
the sphere by a double Wick rotation.
%, and the equations of motion for the $AdS_3$
%$\sigma$-model considered here are very similar to those for the
%$\sigma$-model on $S_3$ discussed in \cite{Kazakov:2004qf}.

%
%\be
%2\pint dy\,\frac{\rho(y)}{x-y}=2\pi n_i-\frac{2\pi J}{\sqrt{\lambda}}
%\left(\frac{1}{x-1}+\frac{1}{x+1}\right),
%\ee
%\be
%\int dx\,\rho(x)=\frac{2\pi}{\sqrt{\lambda}}(\Delta+S-J),
%\ee
%\be
%\int dx\,\frac{\rho(x)}{x}=2\pi m,
%\ee
%\be
%\int dx\,\frac{\rho(x)}{x^2}=\frac{2\pi}{\sqrt{\lambda}}
%(\Delta-S-J).
%\ee
%

%describing the $SU(2)$ sector of the $\CN=4$
%SYM theory, but the roles of the momenta $L,J$ and of the dimension
%$\D$ are played now by $\D-S,\D+S$ and $J$, respectively.

%%%%%%%%%%%%%%%%%%%%%%%%%%%%%%%%%%%%
\subsection{ Comparison of string theory to perturbative gauge theory}
%%%%%%%%%%%%%%%%%%%%%%%%%%%%%%%%%%%%%

 We are now in a position to compare integral equations that encode
periodic solutions of the sigma-model to the scaling limit of Bethe
equations that describe anomalous dimensions in SYM. In order to do
that we need to get rid of the explicit dependence on the angular
momentum $J$ in (\ref{cbethe}). This can be achieved by rescaling
the spectral variable $x$ by
 $4\pi J/\sqrt{\lambda}$:
\be \label{INTEQ}
%G(x+i0)+G(x-i0) =
2\pint dy\,\frac{\rho(y)}{x-y}=2\pi n_i-\frac{x+\frac{m\lambda}{4\pi
J^2}}{x^2 -\frac{\lambda}{16\pi^2J^2}}\,. \ee The normalization
conditions (\ref{nn}), (\ref{anomalous-dim}) and (\ref{cbethemom})
now become
\be
%xG(x)|_{x=\infty}=
\int\label{nnren} dx\,\rho(x)=\frac{S}{J}+\frac{\Delta-S-J}{2J}, \ee
\be\label{mcondSM1}
%-G(0)=
\int dx\,\frac{\rho(x)}{x}=-2\pi m, \ee
\be\label{primG}
%-G'(0)=
\frac{\lambda}{8\pi^2 J}\int dx\,\frac{\rho(x)}{x^2}=\Delta-S-J\,.
\ee
If $\lambda/J^2\rightarrow 0$, we  recover indeed the one-loop Bethe
equations of sec.~\ref{bethesec}. In  section 5 we will obtain the
general solution of this Riemann-Hilbert problem and present
explicitly the one-cut solutions. This solution also demonstrates
the one-loop equivalence of the string and gauge descriptions. It is
interesting that the winding number explicitly enters the right hand
side of the Bethe equation, but it enters in the combination with
the 't~Hooft coupling and therefore disappears at one loop.

The string Bethe equation agrees with the scaling limit of the gauge
Bethe equation up to two loops in the $SU(2)$ sector
\cite{Kazakov:2004qf}. The two-loop agreement extends to the $SO(6)$
(strings moving in $S^5\times \R$), at least for particular
solutions \cite{Minahan:2004ds}. The derivation involves the change
of variables and subsequent expansion in $\lambda /J^2$. Let us try
to  repeat the same steps for the $SL(2)$.  The most important
difference between the string and the gauge Bethe equations is the
normalization of the densities in (\ref{nnren}) and in (\ref{NORM}).
The density for the spin chain (\ref{NORM}) just counts the number
of spin excitations which we would normally identify with the number
of derivatives in the operator (\ref{OPER}). The normalization is
obviously coupling-independent. On the contrary, the normalization
of the string density (\ref{nnren}) depends on the coupling through
$\Delta$. This problem can be fixed by using (\ref{primG}) and
rewriting (\ref{nnren}) as
\begin{equation}\label{}
\int dx\,\rho(x)\left(1-\frac{T}{x^2}\right)=\frac{S}{J}\,,\qquad
\left(T=\frac{\lambda}{16\pi^2J^2}\right).
\end{equation}
The change of variables $x\rightarrow x-T/x$ cancels the unwanted
term in the normalization condition, but spoils the Bethe equation,
since the change of variables, \be\label{nuevar} \pint dy\,
\frac{\rho\left(y+\frac{T}{y}\right)}{x-\frac{T}{x}-y} =\pint
dy\,\frac{\rho(y)}{x-y}+\frac{T}{x}\int\ dy\,\frac{\rho(y)}{y^2}
+\ldots \ee produces a non-local term which explicitly depends on
the density. The other source of non-locality is the winding number
that according to (\ref{mcondSM1}) can also be represented as a
moment of the density. Keeping only $O(\lambda /J^2)$ terms in
(\ref{INTEQ}) we find:
\begin{equation}\label{}
2\pint dy\,\frac{\rho(y)}{x-y}=2\pi
n_i-\frac{1}{x}-\frac{\lambda}{8\pi^2J^2x^3}
-\frac{\lambda}{8\pi^2J^2}\int dy\,\rho
(y)\left(\frac{1}{xy^2}+\frac{1}{yx^2}\right).
\end{equation}
The last term is non-local. Non-localities of this type cancel out
for the $SU(2)$ sigma-model and do not arise in the $SU(2)$ sector
of SYM as well. We cannot exclude that the yet unknown two-loop
corrections make $SL(2)$ Bethe equations non-local\footnote{The fact
that no local modification of the Bethe equations is consistent with
the sigma-model at two loops can be established by analyzing
elliptic (two-cut) solutions \cite{unpublished_Matthias}.} (such
non-local terms might in principle originate from corrections to the
scattering phases of elementary excitations), but it is also
possible that the discrepancies between SYM and strings arise in the
$SL(2)$ sector already at two loops.

%%%%%%%%%%%%%%%%%%%%%%%%%%%%%%%%%%%%%%%%%%%%%%%%%%%%%%%%%%%
%\setcounter{section}{0}
%\appendix
\sectiono{The General Solution of Bethe Equations
in the Scaling Limit \label{ss:kri}}

Here we use the method proposed in \cite{RESHSM} to find the general
solution of (\ref{Bint}). The derivation repeats that for the
compact $SU(2)$ spin chain \cite{Kazakov:2004qf} with minor
modifications.
%We define the function that we shall call
%quasimomentum: \be p(x)=\int dy\,\frac{\rho(y)}{x-y}+\frac{1}{2x}\,.
%\ee
The quasimomentum defined in (\ref{pG}) has a pole at zero and is
analytic elsewhere on the complex plane with cuts $C_i$. The
discontinuity of the quasimomentum across a cut is proportional to
the density and the continuous part is fixed by eq.~(\ref{PPP}). The
function $p(x)$ is completely determined by its analiticity
properties and can be found using the follwing ansatz:
\be\label{DP}
dp=\frac{dx}{y}\left(\frac{a_{-2}}{x^2}+\frac{a_{-1}}{x}+\ldots
+a_{K-2}x^{K-2}\right), \ee
where
\be\label{riemann} y^2=r_0+\ldots r_{2K-1}x^{2K-1}+x^{2K}. \ee
 The quasi-momentum $p(x)$ can now be
obtained by integrating $dp/dx$.

\begin{figure}[tp]
%\hspace*{2cm}
\epsfysize=7cm \centerline{\epsfig{file=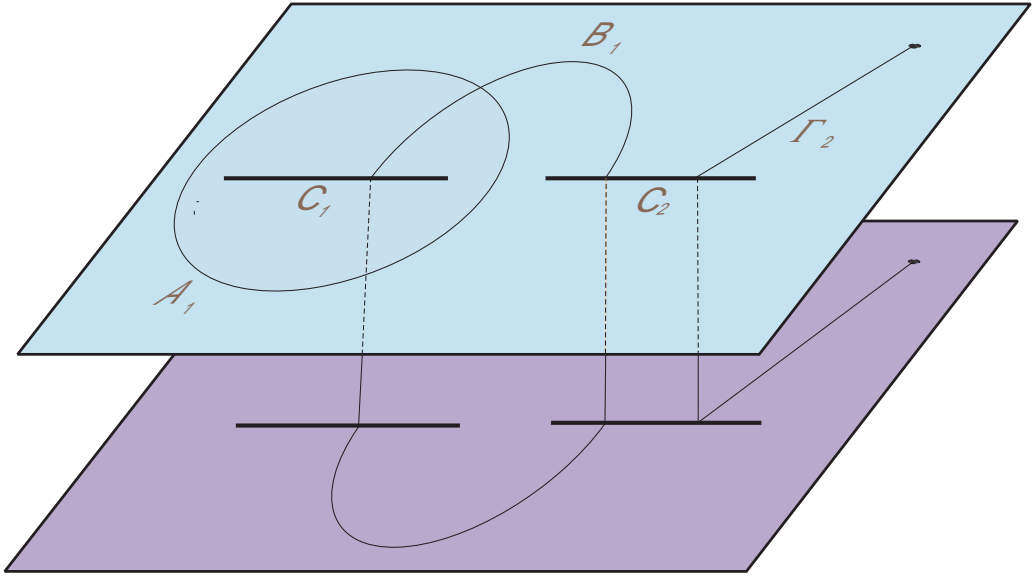,width=140mm}}
\caption{The Riemann surface $\Sigma$ for $K=2$.} \label{riem}
\end{figure}

So defined, $dp$ is an Abelian differential of the third kind on a
hyperelliptic Riemann surface $\Sigma$ of genus $K-1$. The Riemann
surface is defined by (\ref{riemann}). It is obtained by gluing
together two copies of the complex plane along the cuts $C_i$,
$i=1,\ldots,K$. The result of the integration, $p(x)$,  must be
single-valued on the physical sheet. It is easy to see that
single-valuedness of $p(x)$ is equivalent to the vanishing of all
$A$-periods of $dp$: \be\label{Aper} \oint_{A_i}dp=0, \ee where the
$A$-cycles are the contours surrounding the first $K-1$ cuts
(fig.~\ref{riem}).

%%%%%%%%%%%%%%%%%%%%%%%%%%%%%%%%%%%%%%%%%%%
%%%%%%%%%%%%%%%%%%%%%%%%%%%%%%%%%%%%%%%%%

The condition (\ref{PPP}) is equivalent to the integrality of the
B-periods of $dp$: \be\label{Bper} \oint_{B_i}dp=2\pi (n_i-n_K), \ee
where the $B_i$-cycle traverses the $i$th and the $K$th cuts. The
B-cycle conditions constitute $K-1$ linear combinations of the
original $K$ equations (\ref{PPP}). The remaining condition
determines the integral of $dp$ along the open contour $\Gamma_K$
that connects infinite points on the two sheets of the Riemann
surface
% through the $K$th cut $C_K$
(fig.~\ref{riem}): \be\label{Gper} \int_{\Gamma_K}dp=2\pi n_K. \ee

The Laurent expansion of the quasimomentum at zero generates local
charges of the spin chain \cite{Arutyunov:2003rg,Engquist:2003rn}.
In particular: \be p(x)=\frac{1}{2x}-2\pi
m-\frac{8\pi^2J(\Delta-S-J)}{\lambda}\,x+\ldots\,
\qquad(x\rightarrow 0). \ee The condition that $$dp\sim -1/2x^2+{\rm
regular}\qquad(x\rightarrow 0)$$ determines the singular part of
$dp$: \be\label{acoe} a_{-2}=-\frac{\sqrt{r_0}}{2}\,, \qquad
a_{-1}=-\frac{r_1}{4\sqrt{r_0}}\,. \ee The momentum condition is
non-local and requires that \be \int_0^\infty dp\in 2\pi \mathbb{Z}.
\ee The freedom in choosing the contour of integration leads to an
integer-valued ambiguity and thus does not affect the final result.
Finally, the $O(1)$ Laurent coefficient of $dp$ determines the
anomalous dimension: \be \Delta-S-J=\frac{\lambda}{8\pi^2J} \left(
\frac{r_2}{4r_0}-\frac{r_1^2}{16r_0^2}+\frac{a_1}{\sqrt{r_0}}
\right). \ee

Counting the parameters we see that this is the general solution of
the integral equation (\ref{Bint}).  $K$ of the parameters remain
free after imposing the conditions (\ref{Aper}), (\ref{Bper}),
(\ref{Gper}), (\ref{acoe}) on the differential $dp$ and the Riemann
surface $\Sigma$. This $K$-fold ambiguity corresponds to the $K$
filling fractions, the numbers of Bethe roots on each of the cuts
that can be chosen at will. The total number of roots determines the
asymptotics of the quasimomentum at infinity:
$$
p(x)=\left(\frac{S}{J}+\frac{1}{2}\right)\frac{1}{x}+\ldots
$$
and fixes \be a_{K-2}=-\frac{S}{J}-\frac{1}{2}\,. \ee

The simplest solution has only one cut. In that case the
quasimomentum is itself an algebraic function:
\be\label{ONECYM} p(x)=\pi n-\frac{1}{2x}\,\sqrt{(2\pi nx-1)^2-8\pi
mx}\,, \ee
\be\label{SJmn} \frac{S}{J}=\frac{m}{n}\,, \ee
 and
\be\label{DELCOR} \Delta-S-J=\frac{\lambda m(m+n)}{2J}\,. \ee The
anomalous dimension agrees with the energy of the circular string
solution found in
\cite{Arutyunov:2003za}\footnote{Eqs.~(5.30)-(5.32).}. Very similar
solution of the Bethe equations is dual to the pulsating string in
$AdS_3\times S^1$ \cite{Smedback:1998yn}. The solution in
\cite{Smedback:1998yn} describes quite different operators in the
sector with $SO(2,2)$ symmetry, but since $so(2,2)=sl(2)\times
sl(2)$ there are factorized states in the $SO(2,2)$ spin chain for
which the two $sl(2)$'s do not interact. The corresponding solution
of the Bethe equations looks like two copies of the $SL(2)$
solution.

%%%%%%%%%%%%%%%%%%%%%%%%%%%%%%%%%%%%%%%%%%%%%%%%%%%%%%%%%%%
%\setcounter{section}{1}
\sectiono{The General Solution of Sigma-Model Equations
 \label{ss:siggen}}

Here we will find the general finite gap solution of the sigma-model
eq. \rf{INTEQ} and specify it more explicitly for the single cut
case which corresponds to the circular string solution of
\cite{Arutyunov:2003za}.

 We obtained again the same Riemann-Hilbert problem \rf{RHPR},
  as for the long spin chain \rf{PPP},
 but with a different pole structure defined by
 \rf{INTEQ}-\rf{primG} and the definition of the quasimomentum $p(x)$ through the  resolvent
 \rf{spectrepr} with rescaled argument
\be\label{RESREN} G(x)=p(x)-{1\over 4}\({1+4\pi m\sT\over x-\sT}+
 {1-4\pi m\sT\over x+\sT}\)
 \ee
As in sec.~\ref{ss:kri}, we define the differential $dp$  on the
hyper-elliptic surface \rf{riemann}, having double poles
\footnote{We remind that $T={\lambda\over 16\pi^2 J^2}$}
\be\label{polepm} dp\sim dx \left[-{1\mp 4\pi m\sT \over 4( x\pm
\sT)^2}+O\left((x\pm\sT)^0\right)\right]\qquad   {\rm at}\ \ x\to\mp
\sT, \ee
 behaving as
\be\label{poleinf} dp= dx\left[-{\Delta+S\over 2J}{1\over
x^2}+O(1/x^0)\right]\qquad {\rm at}\qquad x\to\infty,  \ee
 and, according to \rf{mcondSM1} and  \rf{primG}
 \be\label{polezero} p(x)= {8\pi^2
J\over \lambda}(S-\Delta)x\qquad {\rm at}\ \  x\to 0  \ee
 We write, generalizing the eq. \rf{DP}
\be\label{twopole} dp={dx\over y}\Big[(1/4+\pi
m\sT)\left({y_+\over (x-\sT)^2}+{y'_+\over x-\sT}\right)  \\
\nonumber  +(1/4-\pi m\sT)\left({y_+\over (x+\sT)^2}+{y'_+\over
x+\sT}\right)+\sum_{k=1}^{K-1} b_k x^{k-1}\Big] \ee
 where $y(x)$ is given by \rf{riemann},
 $y_\pm=\left.y\right|_{x=\pm \sT},\ \ $ $y'_\pm=\left.{dy\over
dx}\right|_{x=\pm \sT}$ due to \rf{polepm}, and the coefficients
$b_k$ and $r_k$ are determined by vanishing of $\bf A$-periods and
fixing the $\bf B$ periods, exactly as in \rf{Aper} and \rf{Gper}.

Let us now find explicitly the single cut solution and   compare it
to the single cut solution of the Bethe equations of the previous
section. The general form of $p(x)$ compatible with the general
solution \rf{twopole} for the chiral field, is \footnote{We keep
here the notations similar to those used for the rational solutions
in \cite{Kazakov:2004qf}}
\be\label{Geq} p(x)=-\frac{1}{4}\left(\frac{(1+\eps)^{-1/2}}{x-\sT}+
\frac{(1-\eps)^{-1/2}}{x+\sT}\right)\sqrt{Ax^2+Bx+C}+\pi n. \ee
 In order to cancel the poles of the "resolvent"
G(x)
 at $x=\pm \sT$ on the physical sheet, we must satisfy the relations
\be\label{bcarel} B=8\pi m+{\eps\over \sT}+16\pi^2
m^2\sT\eps,\qquad C+TA=1+16 \pi^2 m^2 T+8\pi m \sT\eps. \ee
 In order
 to satisfy the momentum condition (\ref{polezero}) we
have
\be\label{crel} p(0)={\sqrt{C}\over
4\sT}\left(\frac{1}{\sqrt{1+\eps}}- \frac{1}{\sqrt{1-\eps}}\right)
+\pi n  =0. \ee
To have the  asymptotic behavior of (\ref{poleinf}) we also require
\be\label{arel}
\sqrt{A}\left(\frac{1}{\sqrt{1+\eps}}+\frac{1}{\sqrt{1-\eps}}\right)
=4\pi n, \ee

 Equations (\ref{bcarel}),
(\ref{crel}) and (\ref{arel}) lead to an equation relating $\eps$
and $J$:
\be\label{kapparel} 1+16\pi^2 m^2T +8\pi m \sT\eps = 16\pi^2
n^2T\frac{1-\eps^2}{\eps^2}.
 \ee
Using the asymptotics  \rf{poleinf} of $p(x)$ we obtain from
(\ref{Geq})  an equation
\be\label{pxinfty}   x p(x)\ \stackreb{ x\to\infty}{\simeq}
-\frac{B}{8\sqrt{A}\sqrt{1-\eps^2}} \(\sqrt{1+\eps}+\sqrt{1-\eps}\)+
\frac{\sqrt{A}\sT}{4\sqrt{1-\eps^2}}
\(\sqrt{1+\eps}-\sqrt{1-\eps}\)=\frac{S+\Delta}{2J}. \ee
%
%Comparing it to \rf{poleinf} we obtain, using  (\ref{bcarel}),
%(\ref{arel}) and \rf{kapparel}, the equation
%
%\be\label{ynorm} \frac{m J}{\sqrt{\lambda}}\eps(1+\sqrt{1-\eps^2})+
%2 n^2\sqrt{1-\eps^2}= \frac{2n(\Delta+S)}{\sqrt{\lambda}}\eps. \ee
%

We also have from \rf{Geq} and \rf{polezero}
\be\label{pprim}  T p'(0)=- \frac{B\sT}{8\sqrt{C}\sqrt{1-\eps^2}}
\(\sqrt{1+\eps}-\sqrt{1-\eps}\)+ \frac{\sqrt{C}}{4\sqrt{1-\eps^2}}
\(\sqrt{1+\eps}+\sqrt{1-\eps}\)=\frac{S-\Delta}{2J}. \ee
%
% Using  \rf{polezero} we get from it the condition
%\be\label{pprim}  \frac{m J}{\sqrt{\lambda}}\eps(1-\sqrt{1-\eps^2})+
%2 n^2\sqrt{1-\eps^2}= \frac{2n(\Delta-S)}{\sqrt{\lambda}}\eps. \ee
%

Equations \rf{kapparel},  and  \rf{pprim}, together with
\rf{bcarel},\rf{crel} and \rf{arel} define the anomalous dimension
$\Delta-S-J$ as a function of $\lambda$ and $J$.

At one loop, our results for the rational solution of this sigma
model match the  corresponding formulas
  for the gauge theory
\rf{ONECYM}-\rf{DELCOR}. For example, in this approximation we
obtain from (\ref{kapparel}) and (\ref{pprim}):
%
%\be   \eps= n\frac{\sqrt{\lambda}}{J}\left(1
%-\frac{n(n+m)^2\lambda}{2 J^2}+O(\lambda^2/J^4)\right), \ee
%
%and inserting it into \rf{pprim} we obtain
%
\be    \frac{\D-S}{J}=1+ \frac{\lambda m(m+n)}{2 J^2}
+O(\lambda^2/J^4), \ee
correctly reproducing the one loop gauge theory formula \rf{DELCOR},
as we expected from the general arguments of the subsection 3.5. The
quasimomentum \rf{ONECYM} can be also easily reproduced in this
approximation from \rf{Geq}-\rf{arel}.

%%%%%%%%%%%%%%%%%%%%%%%%%%%%%%%%%%%%%%%%%%%%%%%%%%%%%%%%%

%%%%%%%%%%%%%%%%%%%%%%%%%%%%%%%%%%%%%%%%%%%%%%%%%%%%%%%%%%%%%%

\sectiono{Discussion
\label{ss:discussion}}

Classical solutions of the sigma-model in the $SL(2)$ sector can be
parameterized by an integral equation of the Bethe type, in complete
analogy with the $SU(2)$ case \cite{Kazakov:2004qf}. These results
may be taken as an indication that the full quantum sigma-model with
\ads\ target (super)space is solvable by some yet unknown quantum
Bethe ansatz. The discretization of the classical Bethe equations
for the $SU(2)$ sector \cite{Arutyunov:2004vx}  reproduces correctly
several quantum effects known from direct calculations. It would be
very interesting to find a discrete counterpart of the Bethe
equations for $SL(2)$ as well. It would be also  interesting to
study the relationship between  the classical limit of the full
one-loop Bethe ansatz in \4N SYM \cite{Beisert:2003yb}  and the full
solution of the classical \ads\ sigma-model which has yet to be
found.

It is generally believed that the weak and strong coupling
calculations of anomalous dimensions agree up to two loops
($O(\lambda^2/J^4)$). Our results may indicate that the
discrepancies occur already at the two-loop level though no
definitive conclusion can be drawn at this point because the $SL(2)$
dilatation operator and the corresponding Bethe equations are not
known beyond one loop.

%%%%%%%%%%%%%%%%%%%%%%%%%%%%%%%%%%%%%%%%%%%%%%%%%%%%%%%%%%%%%%%%%%%

\subsection*{Acknowledgements}

We would like to thank S.~Frolov for drawing our attention to this
problem. We are grateful to N.~Beisert, S.~Frolov, A.~Gorsky,
I.~Kostov, J.~Minahan, M.~Staudacher and A.~Tseytlin for discussions
and comments. V.K. would like to thank IAS (Princeton) for kind
hospitality during the course of this work. K.Z. would like to thank
SPhT, Saclay and AEI, Potsdam for kind hospitality during the course
of this work. The work of V.K. was partially supported by European
Union under the RTN contracts HPRN-CT-2000-00122 and 00131, and by
NATO grant PST.CLG.978817. The work of K.Z. was supported in part by
the Swedish Research Council under contract 621-2002-3920, by
G\"oran Gustafsson Foundation and by RFBR grant NSh-1999.2003.2 for
the support of scientific schools.

\end{document}